\documentclass{astlb}

\usepackage[T2A]{fontenc}
\usepackage[utf8]{inputenc}
\usepackage{lscape}
\usepackage{xspace} 
\usepackage{url}

\usepackage{graphicx}	% Including figure files
\usepackage{amssymb}

\newcommand{\xmm}{XMM-{Newton}}
\newcommand{\ka}{$K\alpha$}

\newcommand{\chan}{\textit{Chandra}}

\def\phflux{ph\,cm$^{-2}$\,s$^{-1}$}

\def\ergscm{erg s$^{-1}$ cm$^{-2}$}

\begin{document}

\journalinfo{(2025)}{51}{8}{19}[29]
%\UDK{524.77}

\title{Direct detection of the non-thermal X-ray emission\\ from the Arches star cluster}

\author{
Roman Krivonos\address{1}\email{krivonos@cosmos.ru},
Alexey Vikhlinin\address{2,1},
Andrei Bykov\address{3},\\
Sergey Sazonov\address{1} and
Ma\"ica Clavel\address{4}
\addresstext{1}{\it Space Research Institute, Russian Academy of Sciences, Moscow, 117997 Russia}
\addresstext{2}{\it Center for Astrophysics $|$ Harvard \& Smithsonian, 60 Garden St, Cambridge, MA, USA}
\addresstext{3}{\it Ioffe Institute, Politekhnicheskaya str., 26, St. Petersburg, 194021, Russia}
\addresstext{4}{\it Univ. Grenoble Alpes, CNRS, IPAG, Grenoble, 38000, France}
}

\shortauthor{Krivonos et al.}

\shorttitle{Non-thermal X-ray emission from the Arches star cluster}

\submitted{28.11.2025 \\
Revised November 28, 2025; Accepted December 1, 2025}

\begin{abstract}
The compact stellar clusters have emerged as particularly promising candidates for cosmic rays (CR) accelerators. The star clusters, recently observed in $\gamma$-rays, are also known sources of non-thermal X-ray emission, which is due to synchrotron or inverse-Compton scattering of relativistic electrons. Thus, the search for the non-thermal X-ray emission from stellar clusters is of particular interest. Until recent time the X-ray emission of the Arches star cluster in the Galactic Center was mixed with non-thermal emission of the surrounding molecular cloud, associated with reflection of hard X-ray irradiation. This reflected emission has been observed to fade, giving us a chance to characterize intrinsic non-thermal emission of the Arches cluster. In this work we demonstrate that Fe K$\alpha$ line emission at 6.4~keV, attributed to the reflected non-thermal emission of the molecular cloud in 2000$-$2010, is not detected in deep observations with {\xmm} in 2020 and {\chan} in  2022, leaving stellar cluster well isolated. We showed that the Arches non-thermal emission is localized in the cluster's core and characterized by a relatively weak, hard ($\Gamma\approx1.5$) power-law spectral continuum with 2$-$10~keV flux of ${\sim}10^{-13}$~\ergscm.


\end{abstract}

\section{Introduction}

The Galactic center (GC) hosts numerous young massive stars, which generate ultraviolet radiation and stellar winds, injecting energy into the interstellar medium (ISM). A significant fraction of these massive early-type stars are hosted in star clusters. There are three main star clusters in the GC: the young nuclear cluster (YNC; Krabbe et al. 1991) which is a part of the Nuclear Stellar Cluster (NSC) but dominates the central parsec, the Arches cluster (Nagata et al. 1995), and the Quintuplet cluster (Nagata et al. 1990). These star clusters are believed to be formed within recent few million years and contain numerous massive stars (Lu et al. 2019).

%There are 13 Wolf-Rayet (WR) stars detected in the Arches cluster and 21 WR stars in the Quintuplet cluster (Espinoza, Selman \& Melnick 2009; Liermann, Hamann \& Oskinova 2012). In addition to the WR stars in the young massive clusters, a comparable number of isolated WR stars are also found distributed throughout the GC (van der Hucht et al. 1996).

When  massive stars reside in binary systems, the intense stellar winds will collide with each other, forming the so-called colliding-wind binaries. The colliding wind generates strong shocks and produces a hot plasma (with a temperature  $>{\sim}10^6$ K). Relativistic particles are also observed in such systems (Eichler \& Usov 1993; Stevens 1995), which can be accelerated through diffusive shock acceleration (Drury 1983). These systems can be bright in the X-ray band. X-ray emission has been detected in several objects including Apep (del Palacio et al. 2023), Eta Carinae (Hamaguchi et al. 2018), and $\gamma$ Vel (Willis, Schild \& Stevens 1995). The overall spectral shape of typical colliding wind sources in X-rays contains a soft thermal component as well as a hard non-thermal component. The latter is believed to originate by synchrotron or inverse-Compton scattering of relativistic electrons, which contributes to hard X-rays and even $\gamma$-rays. The wind velocity and the post-shock temperature determine the thermal component, which contains a bremsstrahlung continuum and emission lines mainly from the K-shell transitions of various elements. 

Detection of very high energy (VHE) $\gamma$-rays from clusters of young massive stars \citep{2019NatAs...3..561A}, provided evidence that acceleration of Galactic Cosmic Rays (CRs) can take place in compact stellar clusters, which can be considered as very high energy CR factories. Particle acceleration by multiple shocks can raise the maximum energy of CR protons out of 1 PeV, making stellar clusters potential candidates for Galactic PeVatrons, especially when the most massive stars produce core-collapsed supernovae \citep[see e.g.][]{2014A&ARv..22...77B}.

Deep observations of rich clusters of young massive stars Westerlund 1 and 2 with {\chan} \citep{2006ApJ...650..203M,2019ApJS..244...28T} and XMM-Newton \citep{2020Ap&SS.365....6K} detected diffuse X-ray emission from the clusters and their close vicinities. Analysis of X-ray spectra of both clusters provided evidence of the possible presence of non-thermal pow-law components \citep{2006ApJ...650..203M,2023MNRAS.525.1553B} that can be produced by synchrotron radiation of very high energy electrons. Evidence for non-thermal X-ray emission was found by \citet{2022A&A...661A..37S} in eROSITA/{\it SRG} spectra of the rich stellar cluster RMC 136 located in the LMC starforming region Doradus 30.  

The young, massive Arches star cluster \citep[G0.12+0.02;][]{1996ApJ...461..750C,1998Natur.394..448S} is located about 25 pc in projection from the Galactic center. The cluster has a compact core that is about $9''$ (${\sim}0.35$~pc at 8 kpc) in radius \citep{1999ApJ...525..750F}, with an average mass density of ${\sim}3 \times 10^{5}$~M$_{\odot}$~pc$^{-3}$, containing more than 160 O-type stars \citep{2002ApJ...581..258F}. There are 13 Wolf-Rayet (WR) stars detected in the  cluster \citep[e.g.,][]{2018A&A...617A..65C,2018A&A...618A...2C,2019A&A...623A..84C}.

%The detection of the continuum radio emission from the Arches cluster suggests both thermal and non-thermal components. \cite{2003ApJ...590L.103Y} argued that high frequency radio emission at 1.4~GHz from the cluster is compact and arises from stellar sources, whereas the low-frequency radio emission at 327~MHz appears diffuse with non-thermal characteristics. 

Based on the first serendipitous {\chan} observations of the Arches cluster region, \cite{2002ApJ...570..665Y} reported the detection of two bright X-ray sources within the central part of the Arches cluster surrounded by diffuse X-ray emission. Authors found that X-ray emission of these sources coincides with the core of the cluster and can be fitted with a two-temperature thermal spectrum with a soft and hard component. The diffuse emission around the cluster was discussed in the context of combined shocked stellar winds escaping from the cluster. 

%Bright X-ray source A1 in the cluster's core was later resolved into two distinct sources \citep{2004ApJ...611..858L,2006MNRAS.371...38W}.

A number of dedicated X-ray campaigns have been initiated to establish the presence of thermal and non-thermal emission components. The thermal emission is thought to originate
from multiple collisions between strong winds of massive stars \citep{1991ApJ...368..241C,2002ApJ...570..665Y,2006MNRAS.371...38W,2011A&A...525L...2C}. The extended, non-thermal X-ray emission has been detected around the cluster \citep{2006MNRAS.371...38W,2007PASJ...59S.229T,2011A&A...530A..38C,2012A&A...546A..88T,2014ApJ...781..107K}. The non-thermal origin of diffuse emission was revealed by fluorescent Fe {\ka} line emission of neutral or low ionized material. The fluorescent emission is produced by ejecting a K-shell electron, either by hard X-ray photoionization from an external X-ray source (reflection mechanism), or by the collisional ionization induced by cosmic-ray (CR)-accelerated electrons or ions (CR-induced  scenario). 

The mechanism producing the fluorescent line emission around the Arches cluster was not completely understood for a long time. Based on {\xmm} data, \citet[][hereafter T12]{2012A&A...546A..88T} argued that bright 6.4 keV line extended emission observed around the Arches cluster is very likely produced by bombardment of molecular gas by energetic ions. This statement was mainly based on apparently constant X-ray flux, however, \cite{Clavel2014} analyzed the long-term behavior of the Arches cluster molecular cloud (or simply the Arches cloud) over 13~yr with {\xmm} and reported a 30\% decrease in Fe {\ka} line and continuum flux of the cloud emission, providing significant evidence for the X-ray reflection scenario. Later, \cite{2017MNRAS.468.2822K}, with long NuSTAR and {\xmm} observations in 2015 showed that the non-thermal emission of the Arches cloud vanished after 2012, leaving three bright clumps. The declining trend of the cloud emission was consistent with half-life decay time of ${\sim}8$~yr, similar to several other molecular clouds in the Galactic Center.

\begin{table}
\centering
\caption{Log of {\xmm} observations of the Arches cluster, used in this work.}
\label{tab:log:xmm}
\begin{tabular}{cccc}
\hline
Date & ObsID & Exp., ks & Offset, $'$\\
\hline
2020-09-19 & 0862470701 & 43 & 4.14 \\
2020-09-21 & 0862470601 & 43 & 6.36 \\  
2020-09-26 & 0862470801 & 43 & 10.37 \\

%2021-03-28 & 0860620201 & 28 & 11.24 \\
%2021-03-29 & 0860620301 & 31 & 11.24 \\
%2021-03-30 & 0860620401 & 31 & 11.24 \\

%2022-03-18 & 0893811101 & 54 & 11.24 \\
%2022-03-26 & 0893811301 & 56 & 11.24 \\

%2024-03-28 & 0944580801 & 26 & 11.24 \\
%2024-03-29 & 0944580501 & 26 & 11.24 \\
\hline
\end{tabular}
%\begin{flushleft}
%$^{\rm a)}$ 
%\end{flushleft}
\end{table}

Thus, the study of the intrinsic non-thermal emission of the Arches star cluster was complicated by presence of the nearby cloud emission. The detection of 6.4 keV line within the cluster points out to this `contamination', e.g. T12 suggested that 6.4 keV line in the cluster's core is produced in molecular gas along the line of sight outside the star cluster. Therefore, the long-term fading of the Arches cloud \citep{Clavel2014,2019MNRAS.484.1627K,2025A&A...695A..52S}, opens a possibility to observe intrinsic non-thermal emission of the cluster. The non-detection of 6.4 keV line in the cluster would indicate that possible contribution of the reflected emission is not significant.

%With NuSTAR observations in the hard X-ray domain, \cite{2014ApJ...781..107K} showed that the spatial distribution of the diffuse emission 10$-$20~keV is consistent with the broad region around the cluster where the 6.4 keV line is observed.

In this work we analyze deep and moderately off-axis {\xmm} observations of the Arches cluster taken in 2020 during Central Molecular Zone (CMZ) survey, with the aim to characterize its intrinsic non-thermal emission, having the fact that the reflected X-ray emission of the surrounding molecular cloud is not significantly detected. Additionally, we utilize {\chan} observations of the Arches cluster, taken in 2022 as a part of deep survey of CMZ region \citep{2023Natur.619...41M}. The paper is organized as follows. Sect.~\ref{sec:data} describes the observations and data reduction, including sky map production and extraction of X-ray spectra. In Sect.~\ref{sec:res} we present results for morphology analysis and spectral decomposition of the Arches X-ray emission. Discussion, summary and conclusions are given in  Sections~\ref{sec:discussion}-\ref{sec:summary}.

%# Какие наклоны разрешены для какой температуры?
%# W1 жесткость от центра? Муно и др.
%# есть ли двойные системы? Кларк и др.

\section{Observations and data reduction}
\label{sec:data}

We analyzed deep observations of the Arches cluster region, taken during observational campaigns of CMZ with {\xmm} in 2020 (PI: Clavel) and {\chan} in 2022 (PI: Vikhlinin). We selected only observations with Arches cluster appeared in the field of views's (FOV's) of both observatories. Table~\ref{tab:log:xmm} contains the list of used {\xmm} observations in 2020. 

%The {\xmm} data reduction was carried out using the {\xmm} Extended Source Analysis Software \citep[\texttt{ESAS},][]{snowden2004,snowden2008}, included in version 22 of {\xmm} Science Analysis Software (\texttt{SAS}).

\begin{figure}
	\includegraphics[width=\columnwidth]{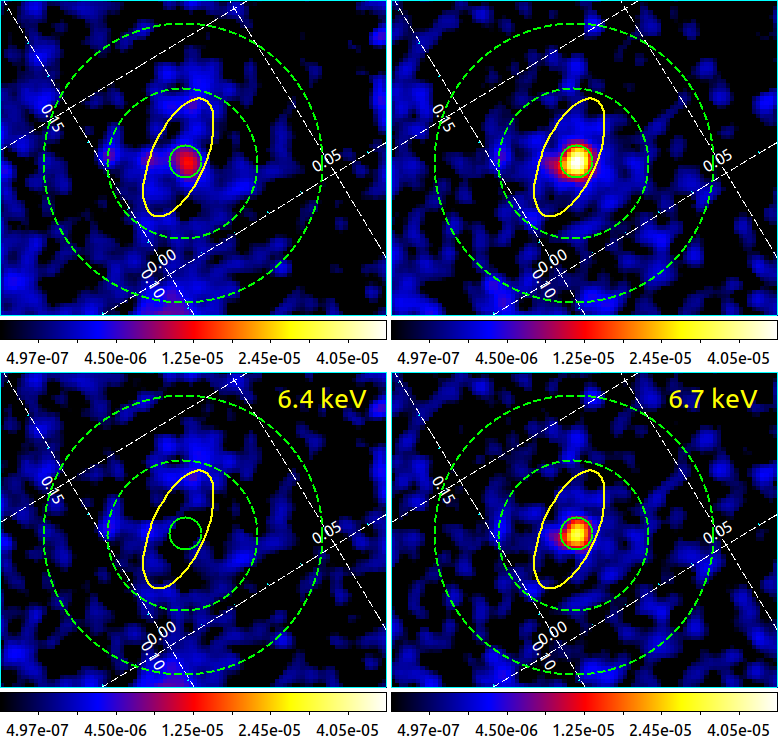}
    \caption{{\it Upper panels:} {\xmm}/EPIC exposure$-$ and background$-$corrected maps of the Arches cluster region containing the line (and continuum) emission at 6.4 keV ({\it left panels}) and 6.7~keV ({\it right panels}) observed in 2020. The images, initially corrected for particle background, were corrected for astrophysical background measured in the green dashed annulus. {\it Bottom panels:} sky images, additionally corrected for continuum emission, represent Fe K$\alpha$ emission line maps at 6.4 keV and 6.7 keV. The green circle with radius of $15''$ indicates the region used to characterize the Arches cluster X-ray emission, which shows strong Fe K$\alpha$ emission at 6.7~keV. Yellow ellipse denotes reference region of the Arches molecular cloud observed earlier at 6.4~keV, adopted from T12. Green dashed annulus is also used in spectral analysis to extract background. All sky images were convolved with a Gaussian ($\sigma=1.5$~pix, the pixel size is $4''$).}
    \label{fig:lines:map}
\end{figure}

\subsection{{\xmm} observations in 2020}
\label{sec:xmm}

The {\xmm} data reduction was carried out using the {\xmm} Science Analysis Software (\texttt{SAS}) version 22. Calibrated event files were produced using the tasks \texttt{emproc} for the MOS cameras and \texttt{epproc} for the pn camera. We excluded from the analysis the period contaminated by soft proton flares using \texttt{espfilt} from the Extended Source Analysis Software \citep[\texttt{ESAS},][]{snowden2004,snowden2008} package, included in SAS. We selected events with \texttt{XMMEA\_EM\&PATTERN} $\leq$~12 and \texttt{XMMEA\_EP\&PATTERN} $\leq$~4 for MOS cameras and pn instrument, respectively.  The particle background was derived from Filter-Wheel Closed (FWC) observations using \texttt{evqpb} task from ESAS. We applied exactly the same event selection criteria to both data and FWC files (see  also Sect.~\ref{sec:morphology}).

% T12 used XMMEA_EM and XMMEA_EP for imaging and XMMEA_SM (MOS) FLAG=0 (pn) for spectra

Similar to T12, we produced count images in three energy bands 4.17$-$5.86~keV, 6.3$-$6.48~keV and 6.564$-$6.753~keV, which contain the continuum emission, the Fe K$\alpha$ lines at 6.4~keV from neutral to low-ionized atoms and at 6.7~keV from a hot thermally-ionized plasma. For each observation, instrument, energy band, the normalized FWC image was subtracted from the count image. The corresponding vignetting-corrected exposure maps were generated with \texttt{eexpmap} task.

The Arches cluster core emission is embedded in the elongated non-thermal emission of the cloud with dimensions of ${\sim}25 {\times} 59$~arcsec$^2$, which approximately corresponds to yellow ellipse in Fig.~\ref{fig:lines:map}. In order to perform spatial and spectral analysis consistent with other studies of the Arches cluster we adopt the same sky regions to describe the core of the Arches cluster and the surrounding cloud region.

We extracted MOS1, MOS2 and pn spectra for each observation from the Arches cluster core region, described as a circle with radius of $15''$, which is selected to cover the compact cluster's core taking {\xmm} point spread function into account. It also corresponds to the same Arches cluster core region used in T12 and other works \citep{Clavel2014,2014ApJ...781..107K,2017MNRAS.468.2822K,2019MNRAS.484.1627K}.  The background spectrum was measured in annulus region with internal and outer radius of $70''$ and $130''$, respectively.

Spectral analysis was done using \textsc{XSPEC} \citep{xspec}, setting the atomic cross sections to \cite{1996ApJ...465..487V} and the abundances to \citep{2000ApJ...542..914W}. We merged MOS1, MOS2 and pn spectra for each observation using \texttt{epicspeccombine} task.

\subsection{{\chan} observations in 2022}
\label{sec:chandra}

We use {\chan} data from deep observations of the CMZ region performed in 2022 with a total exposure of 900~ks. A part of this survey also covers the Arches cluster zone. The detailed description of {\chan} observations and the procedure for data analysis can be found in \cite{2023Natur.619...41M}; see also \citet{2009ApJ...692.1033V}. For better modeling of the instrumental background, the data have been organized in three epochs 2022-03, 2022-07 and 2022-09. Spectral extraction between 1 and 9 keV of the Arches cluster emission has been made from the same circular region with $15''$ radius, used throughout this work.

\section{Results}
\label{sec:res}

\subsection{Morphology}
\label{sec:morphology}

In Fig.~\ref{fig:lines:map} (upper panel) we show exposure$-$ and FWC$-$ corrected maps of the Arches cluster region in the energy bands containing 6.4 keV and 6.7~keV line emission as observed with {\xmm} in 2020. The images were additionally subtracted with mean astrophysical background measured in the annulus region. However, in order to produce sky maps of the Fe {\ka} emission at 6.4 and 6.7 keV, the underlying continuum needs to be subtracted. To this end, the continuum-dominated 4.17$-$5.86~keV map was renormalized to the power-law flux in the considered energy band and then subtracted from the corresponding energy band image. We use absorbed power-law model with photon index $\Gamma=3$, consistent with spectral analysis presented below. The resulting continuum-corrected sky images of the Arches complex (Fig.~\ref{fig:lines:map}, bottom) demonstrate that 6.4~keV Fe line structures observed around the Arches cluster (inside the reference ellipse region) in 2000-2010 and attributed to the molecular cloud emission are not detected in 2020, including the core of the cluster. This fact is also confirmed by strong suppression of Fe K$\alpha$ emission of the cluster region observed in multi-year {\xmm} maps of the CMZ \citep{2025A&A...695A..52S}. Bright 6.7~keV line emission, as expected, is concentrated in the cluster's core.

\begin{figure}
\includegraphics[width=\columnwidth]{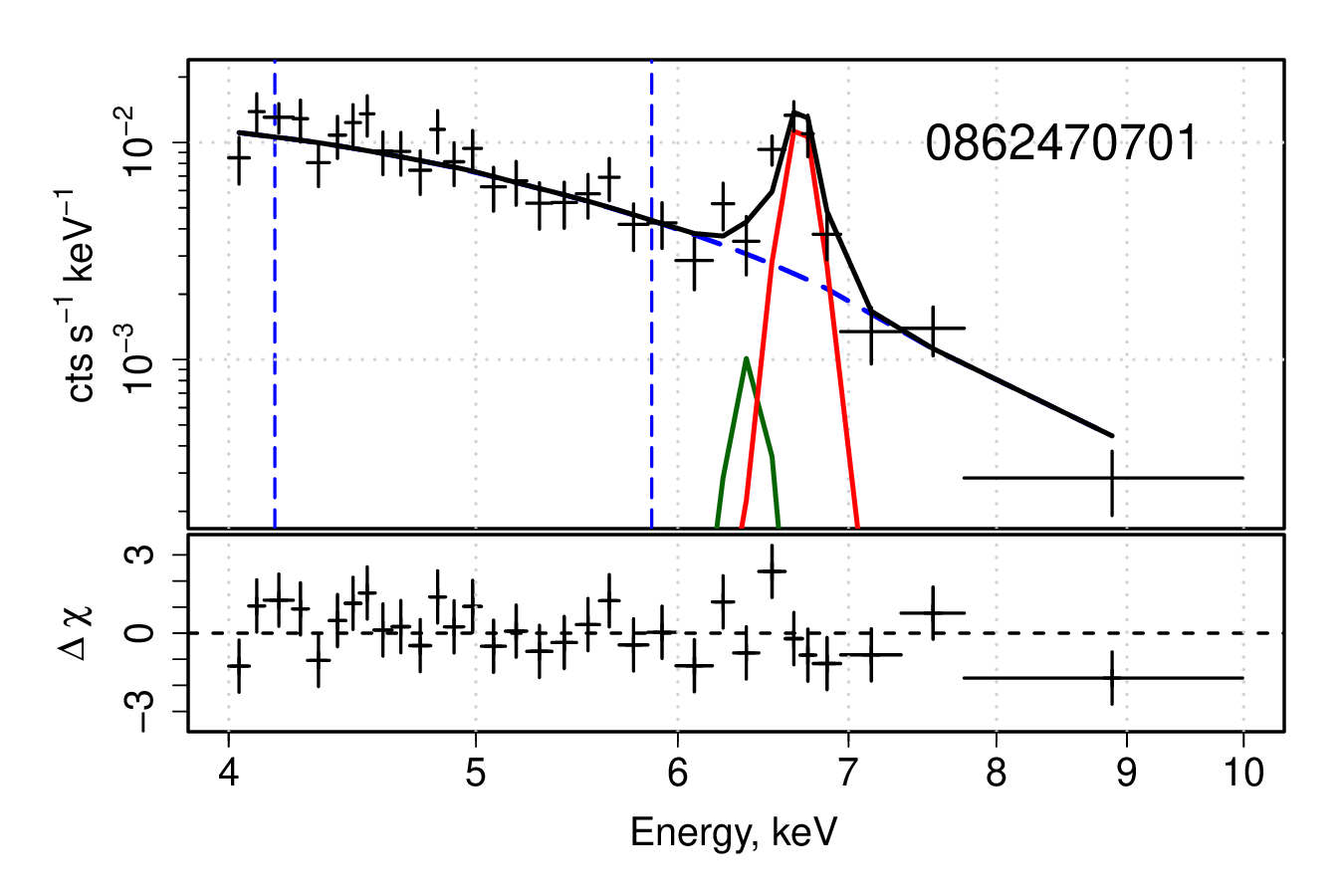}
\includegraphics[width=\columnwidth]{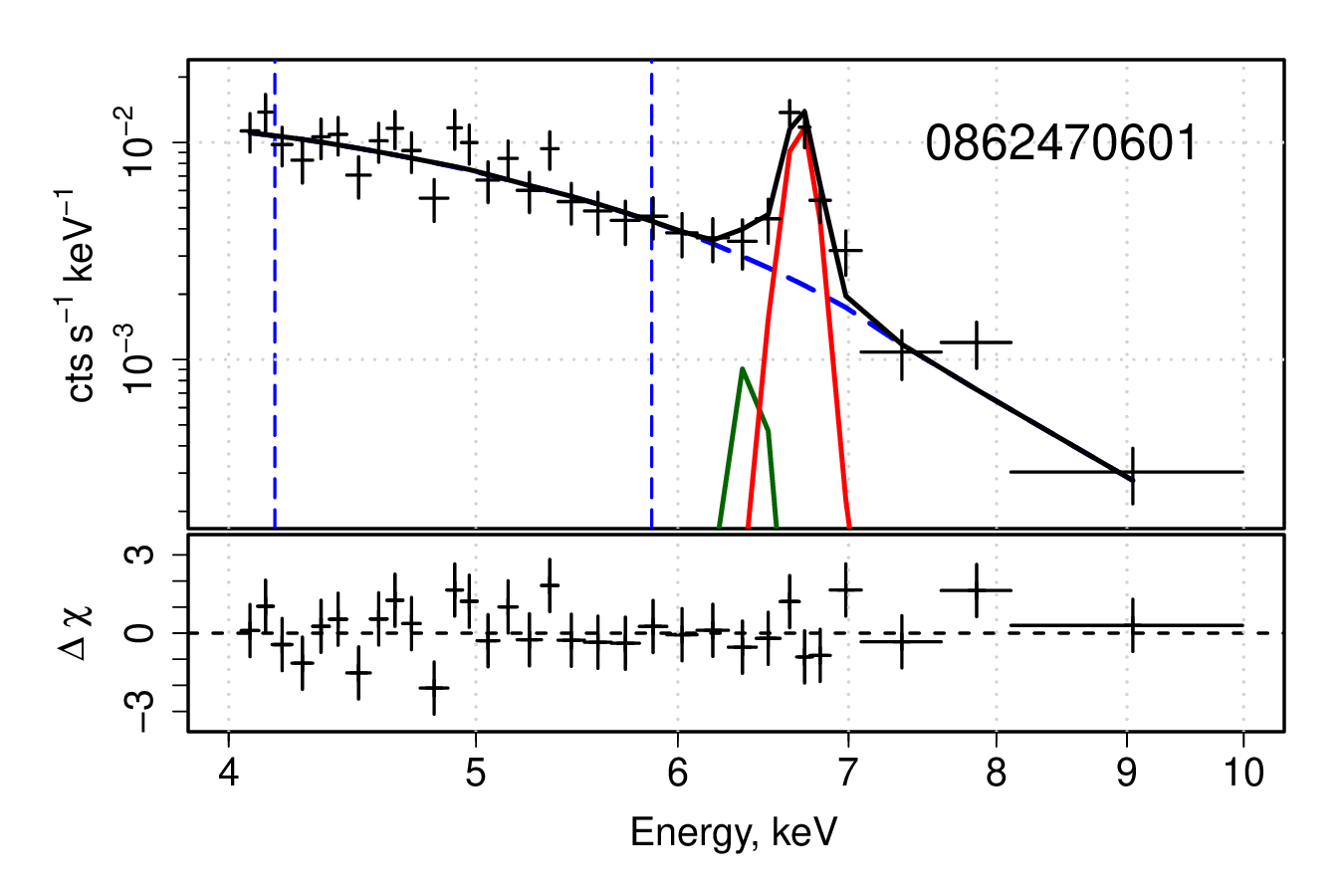}
\includegraphics[width=\columnwidth]{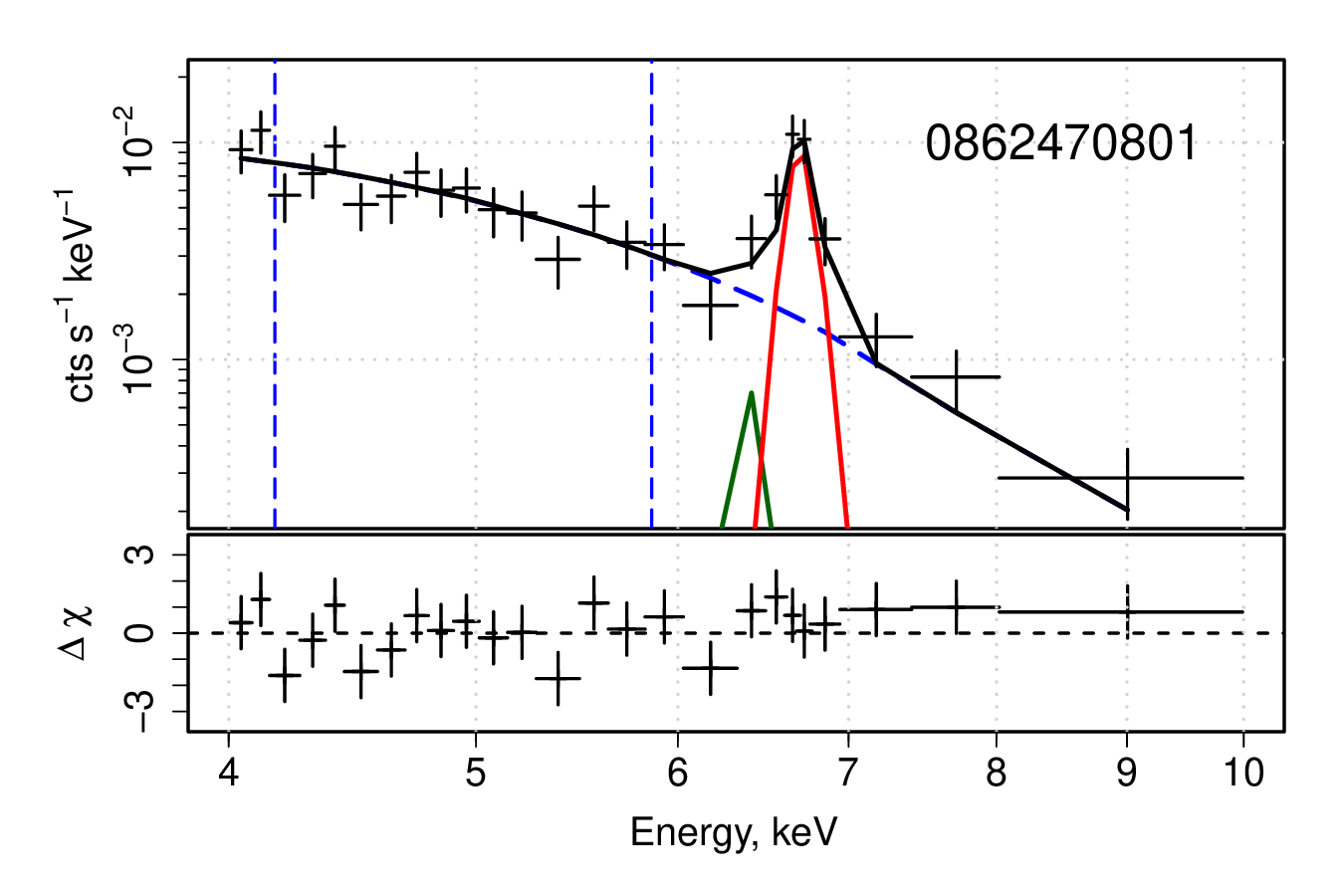}
    \caption{Merged MOS1, MOS2 and pn X-ray spectra of the Arches cluster between 4 and 10 keV for three {\xmm} observations in 2020. Best-fitting model (solid black line) is composed by a combination of an absorbed power-law (dashed blue) normalized in energy range shown by vertical blue dashed lines, and two Gaussian functions attributed to Fe K$\alpha$ lines at 6.4 keV (dark green) and 6.7 keV (red).}
    \label{fig:lines:spe:xmm}
\end{figure}

\begin{figure}
\includegraphics[width=\columnwidth]{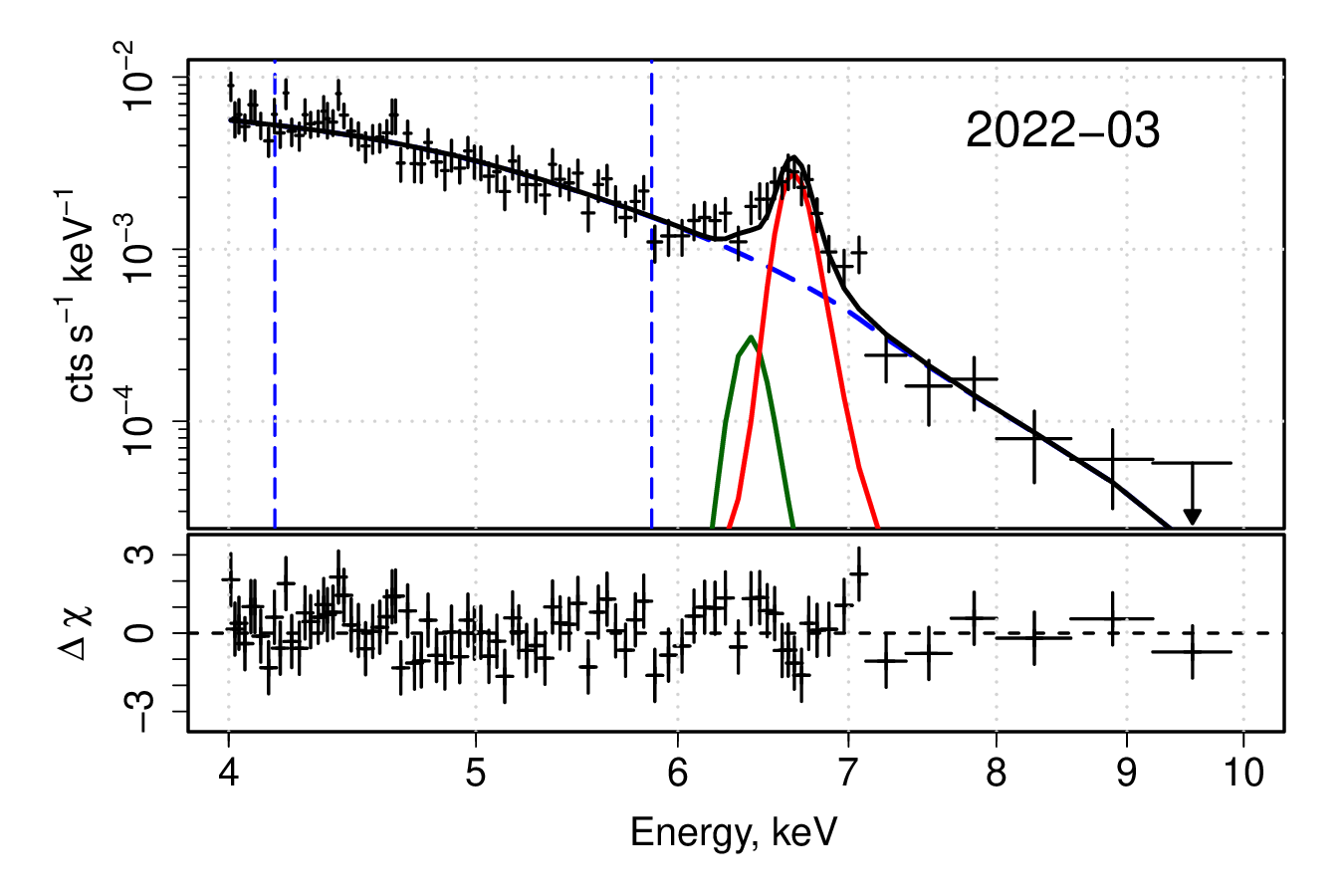}
\includegraphics[width=\columnwidth]{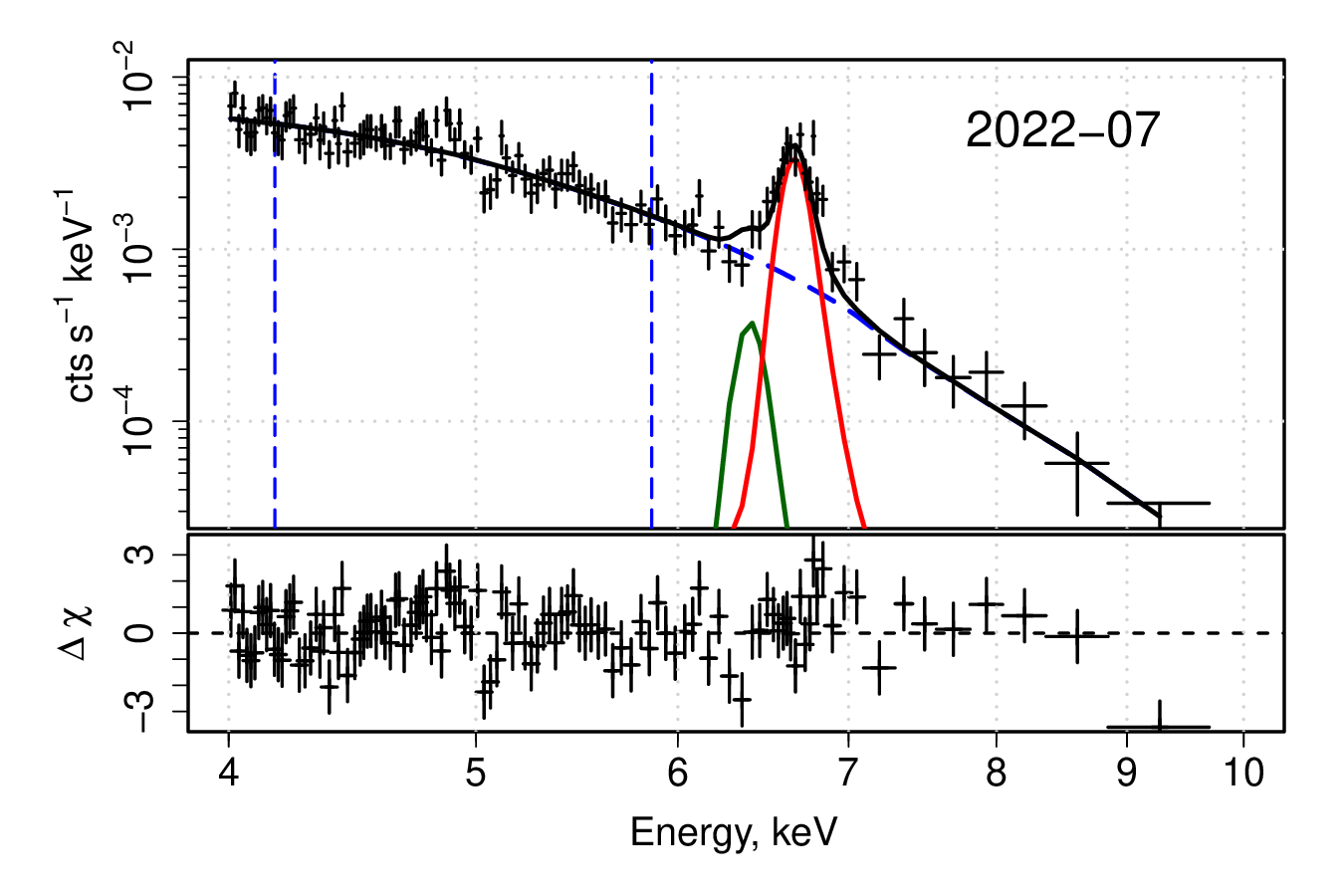}
\includegraphics[width=\columnwidth]{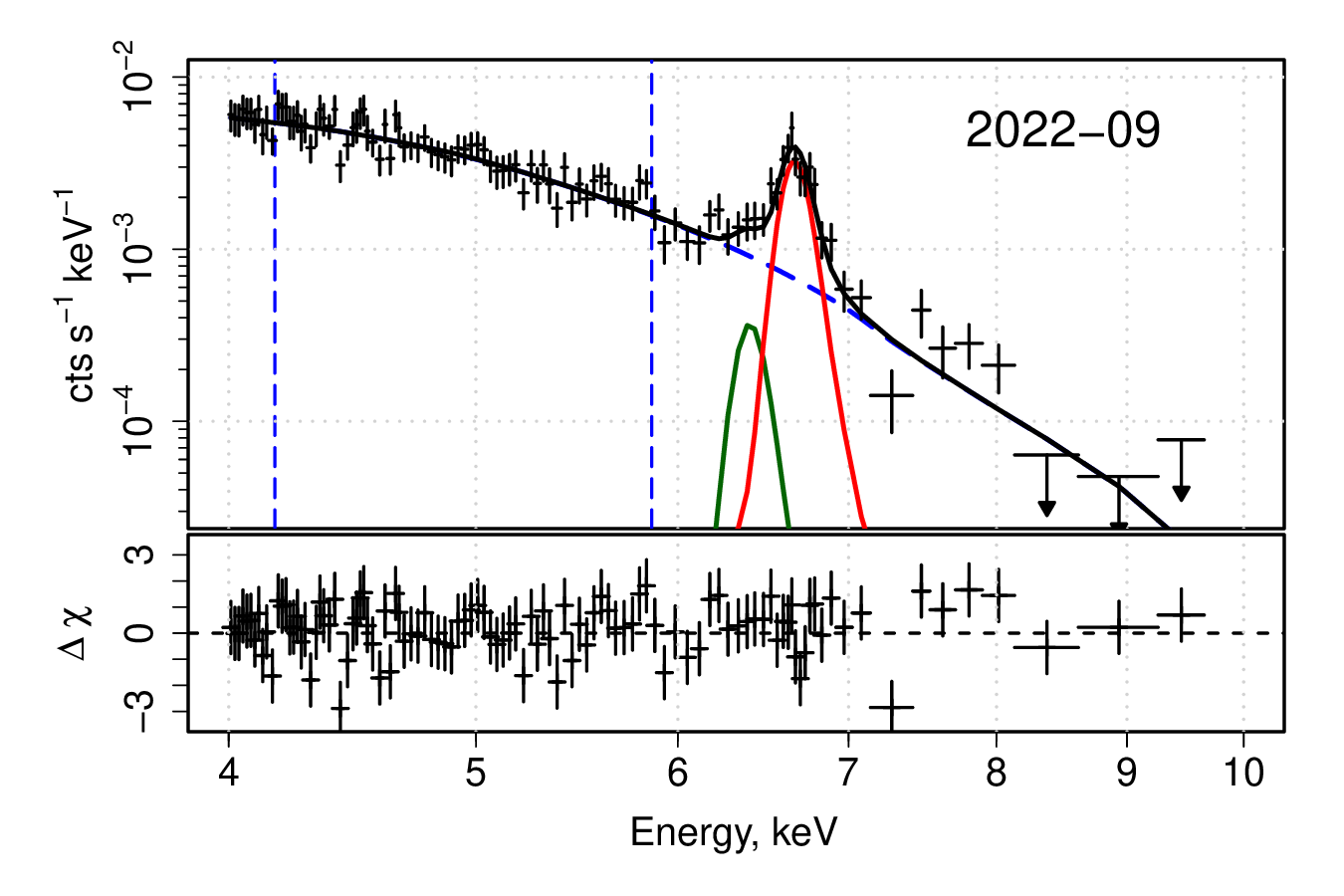}
    \caption{Chandra X-ray spectra of the Arches cluster between 4 and 10 keV for three epochs in 2022. See Fig.~\ref{fig:lines:spe:xmm} for reference.} 
    \label{fig:lines:spe:chandra}
\end{figure}

\subsection{Spectral analysis}

\begin{table*}
\centering 
\caption{Best-fitting parameters of a power-law model (Eq.~\ref{eq:pow}). }
\label{tab:spec:pow}
\begin{tabular}{llllll}
\hline
 &  & \multicolumn{2}{c}{{\xmm} 2020} & \multicolumn{2}{c}{{\chan} 2022} \\ 
Parameter & Units & ObsID & Value & Epoch & Value \\ 
%\hline
%Sys. error  & 5\%  & 0\% & 0\% & 15\% \\
\hline
$const$ & & 0862470701 & 1.0 (fixed) &2022-03& 1.0 (fixed)\\ 
$const$ & & 0862470601 & $1.18\pm0.11$ &2022-07&$1.01\pm0.05$\\ 
$const$ & & 0862470801 & $1.06\pm0.11$ &2022-09&$1.02\pm0.05$\\ 
$N_{\rm H}$  & $10^{22}$~cm$^{-2}$ & & 9.5 (fixed) &&9.5 (fixed)\\ 
$\Gamma^{\rm pow}$ & & & $3.02\pm0.25$ &&$3.22\pm0.15$\\ 
$F_{\rm 4.17-5.86~keV}$ & $10^{-13}$~\ergscm & & $1.71\pm0.12$ &&$2.16\pm0.08$\\ 
$E_{\rm 6.4~keV}$ & keV & & 6.4 (fixed) &&6.4 (fixed)\\  
$F_{\rm 6.4~keV}$ & $10^{-7}$~\phflux & & $<8$ &&$5.3\pm2.8$\\  
EW$_{\rm 6.4~keV}$ & keV & & $<0.1$ &&$0.11\pm0.02$\\

$E_{\rm 6.7~keV}$ & keV & & $6.677\pm0.015$  &&$6.663\pm0.011$\\  
$F_{\rm 6.7~keV}$ & $10^{-6}$~\phflux & & $5.3\pm0.7$ &&$5.7\pm0.5$\\  
EW$_{\rm 6.7~keV}$ & keV & & $0.98\pm0.08$ &&$0.90\pm0.05$\\
$\chi^{2}_{\rm red}$/dof & & & 0.93/82 &&1.06/305\\ 
\hline
\end{tabular}
\end{table*}

In the Arches cluster region T12 identified an optically thin thermal plasma with a temperature $kT$ $\sim$ 1.6$-$1.8~keV and a relatively weak non-thermal component characterized by a hard continuum emission and a line at 6.4~keV from neutral to low-ionized Fe atoms (EW$_{\rm 6.4~keV}$ = $0.4 \pm 0.1$~keV). T12 suggested that the 6.4 keV line detected from this region is produced in molecular gas along the line of sight outside the star cluster. To test this assumption, we start with constraining non-thermal molecular cloud emission traced by Fe K$\alpha$ 6.4 keV line in the projected region of the Arches cluster core. To this end, we simplify spectral fitting by limiting energy range to 4$-$10~keV and considering 6.4~keV and 6.7~keV line emission above the power-law continuum with free photon index and normalization. We used the following model:

\begin{equation}
\label{eq:pow}
const \times ( wabs \times pegpwrlw + gaussian + gaussian )
\end{equation}

where $const$ term describes cross-normalization between different observations; $wabs$ is an absorption term, chosen to be consistent with previous works, with fixed $N_{\rm H} = 9.5 \times 10^{22}$ cm$^{-2}$ according to T12; $pegpwrlw$ -- a power-law with normalization in 4.17$-$5.86~keV band and free photon index $\Gamma$ linked between all observations; first Gaussian is 6.4 keV line with fixed energy 6.4~keV; second Gaussian describes 6.7 keV line with free energy. Both Gaussian components have width fixed at $\sigma=10$~eV. Note that we do not correct the iron line emission for the overall absorption \citep[see][for explanation]{Clavel2014}. The {\xmm} 2020 and {\chan} 2022 spectra were fitted separately. The best fitting parameters are listed in Table~\ref{tab:spec:pow} and spectra are shown in Figs.~\ref{fig:lines:spe:xmm}-\ref{fig:lines:spe:chandra}. The equivalent width (EW) of iron lines have been estimated with \textsc{SHERPA} modeling and fitting package \citep{2001SPIE.4477...76F,2024ApJS..274...43S}, a part of the \textsc{CIAO} software \citep{ciao}. In {\xmm} 2020 spectrum, the EW of 6.7~keV line was found at the level of ${\sim}1$~keV, while 6.4~keV line has not been significantly detected, and fitting procedure provided upper limit EW$_{\rm 6.4~keV}<0.1$~keV (68\% confidence level). Taking the last into account, we removed 6.4 keV line from the following {\xmm} spectral analysis. The {\chan} 2022 spectrum shows marginal detection of 6.4~keV line with EW$_{\rm 6.4~keV}\approx0.1$~keV, consistent with {\xmm} upper limit. Note that the spectral analysis of both 2020 and 2022 data sets demonstrate consistent model parameters.

\begin{figure}
\includegraphics[width=\columnwidth]{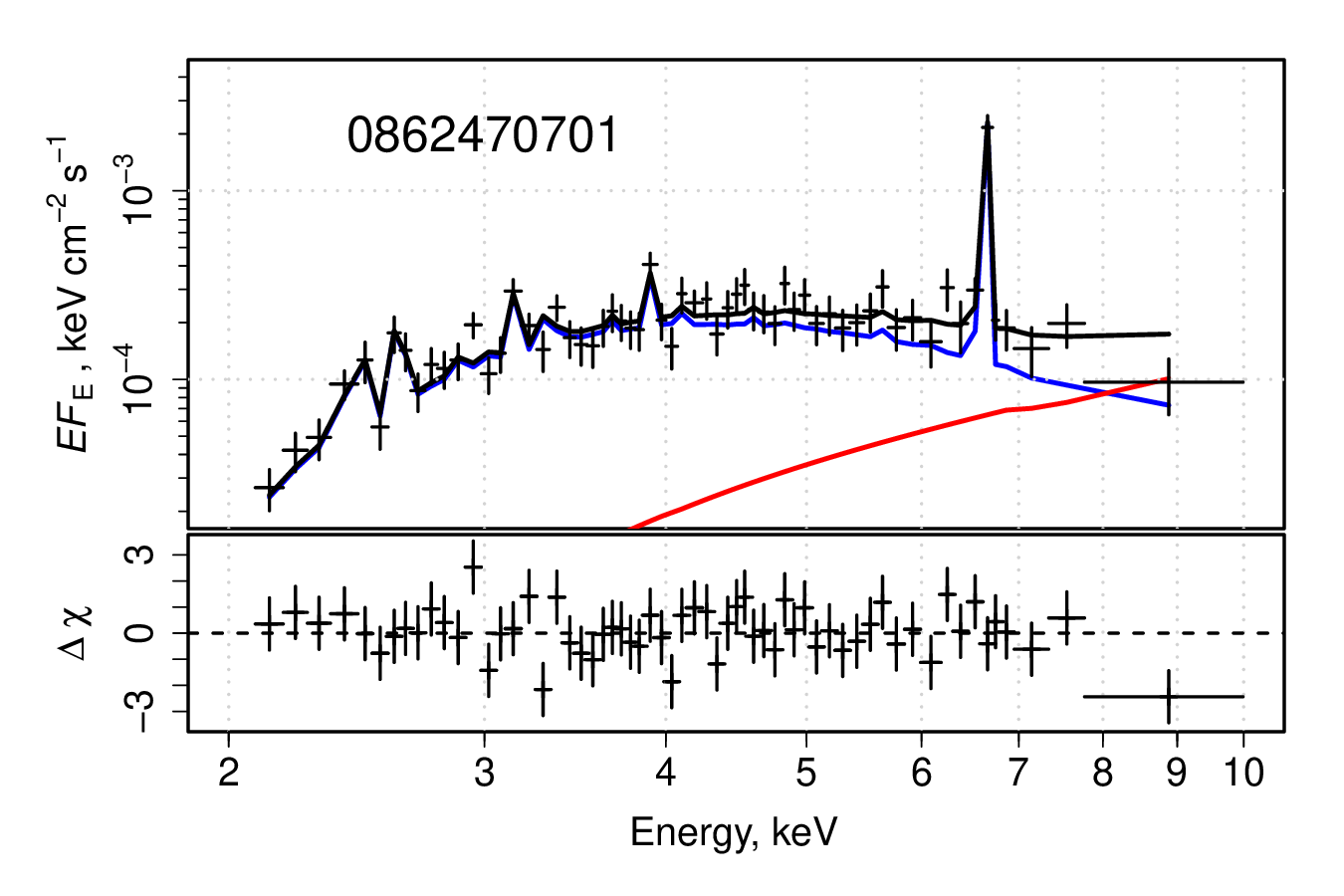}
\includegraphics[width=\columnwidth]{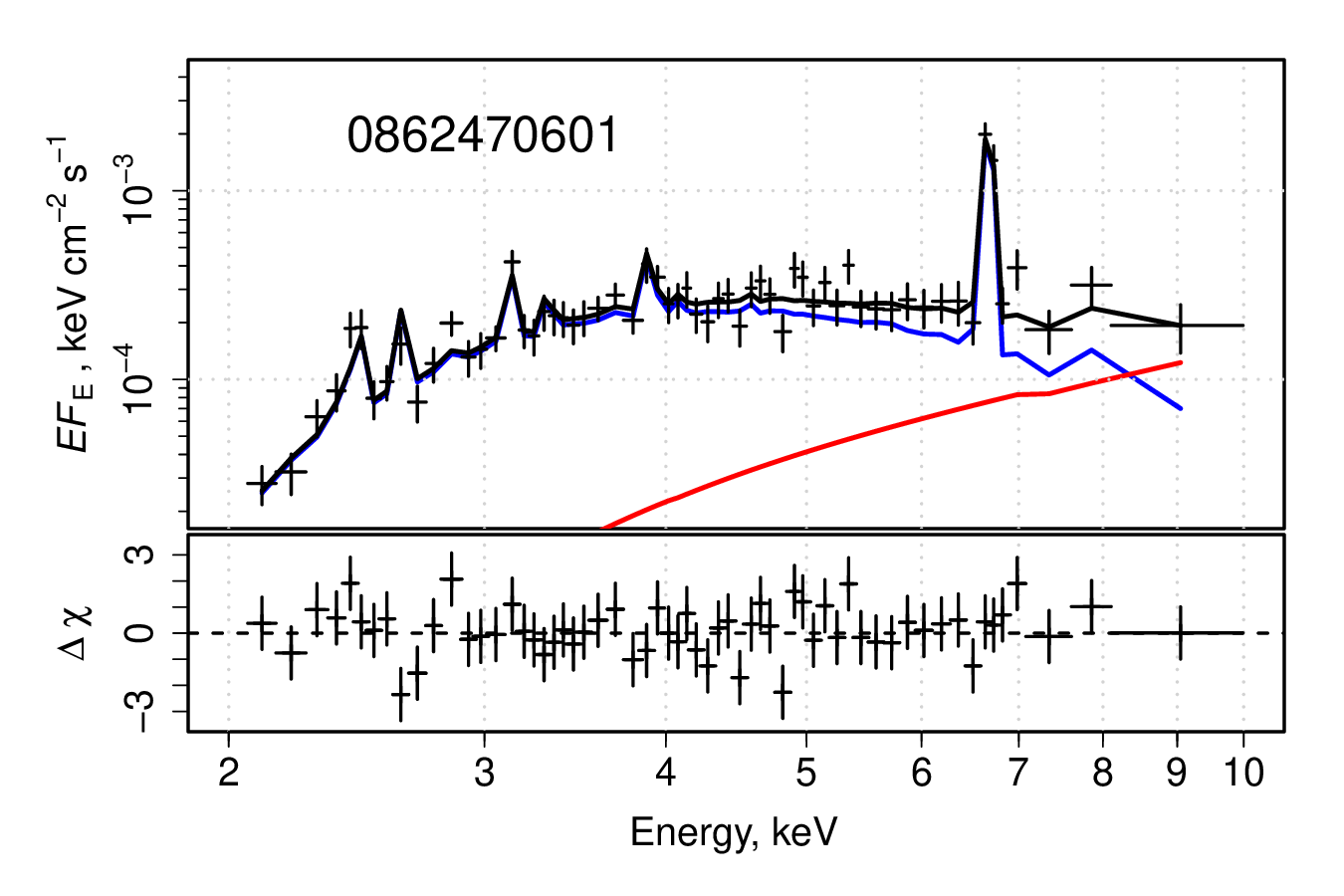}
\includegraphics[width=\columnwidth]{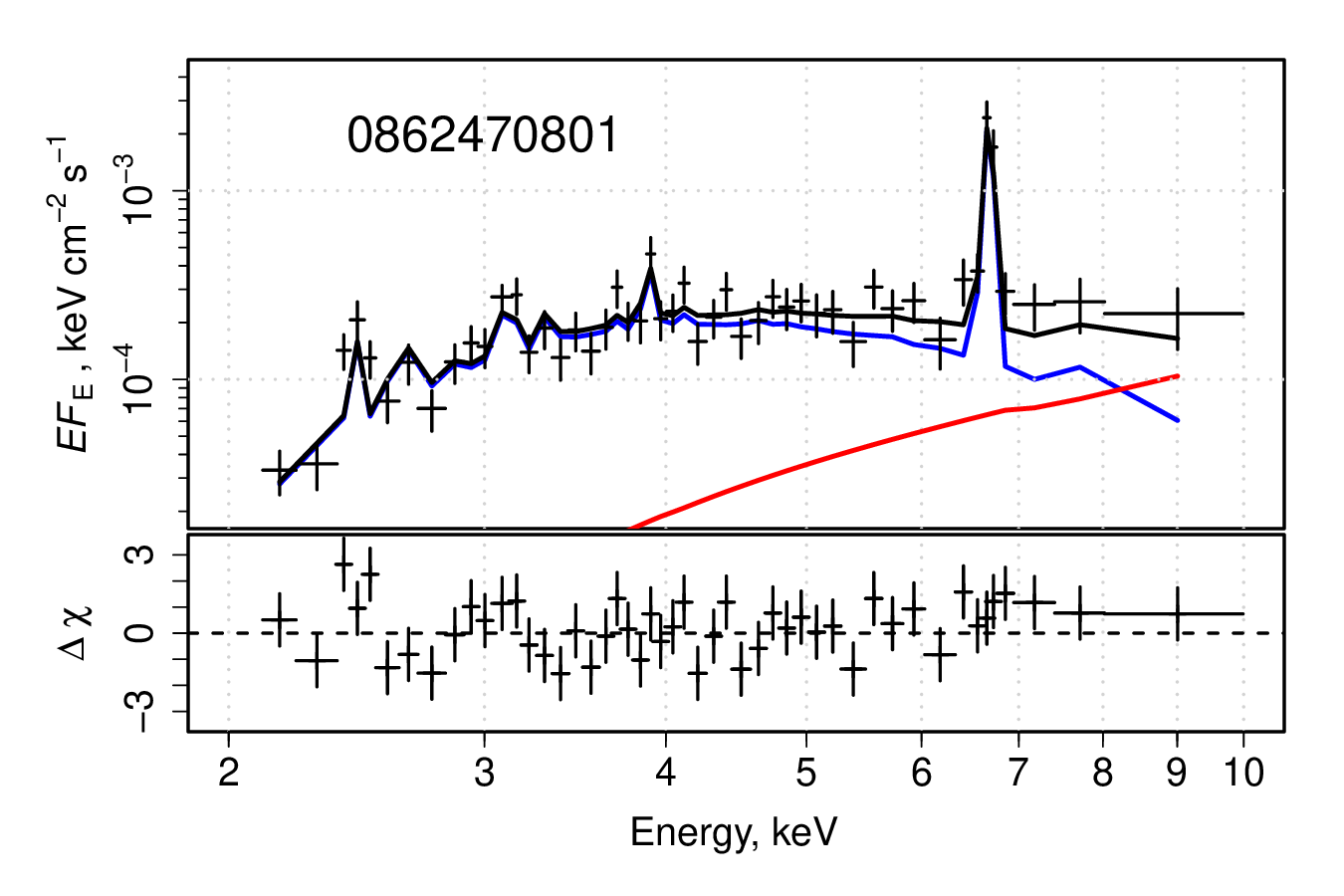}
    \caption{Merged MOS1, MOS2 and pn X-ray spectra of the Arches cluster between 2 and 10 keV for three {\xmm} observations in 2020. Best-fitting model (solid black line) is composed by a combination of an absorbed \texttt{apec} (blue) and power-law (red).}
    \label{fig:apec:spe:xmm}
\end{figure}

\begin{figure}
\includegraphics[width=\columnwidth]{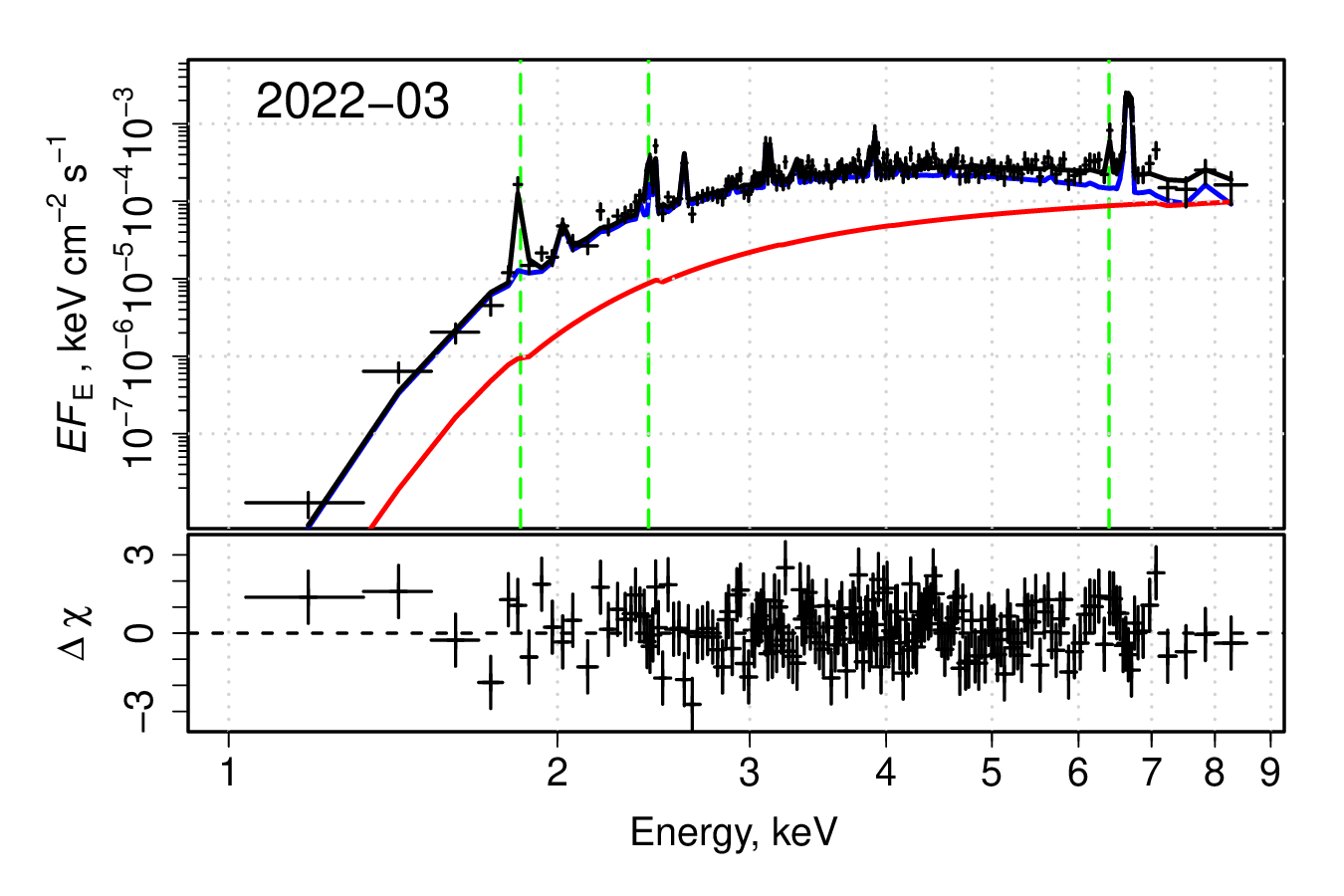}
\includegraphics[width=\columnwidth]{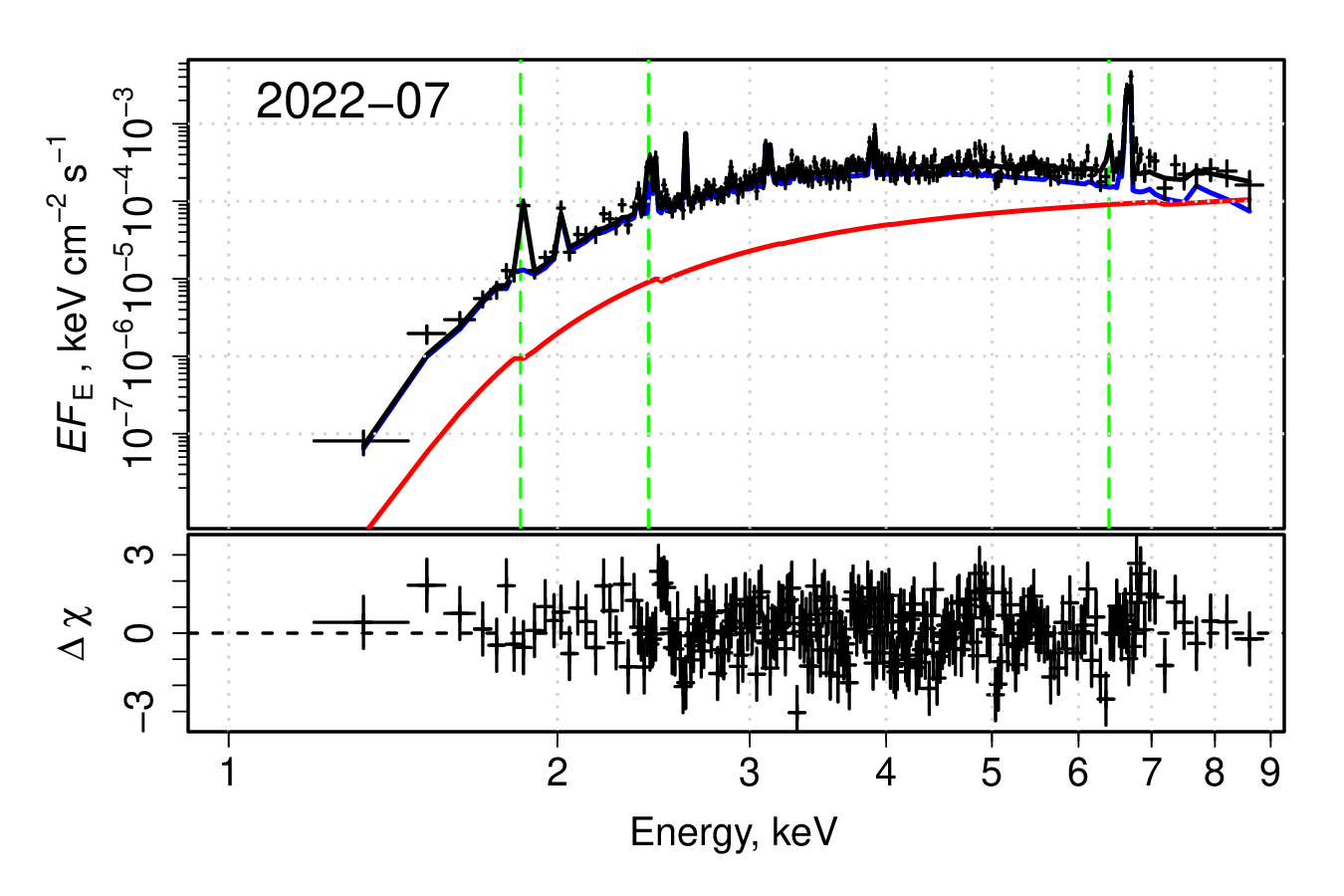}
\includegraphics[width=\columnwidth]{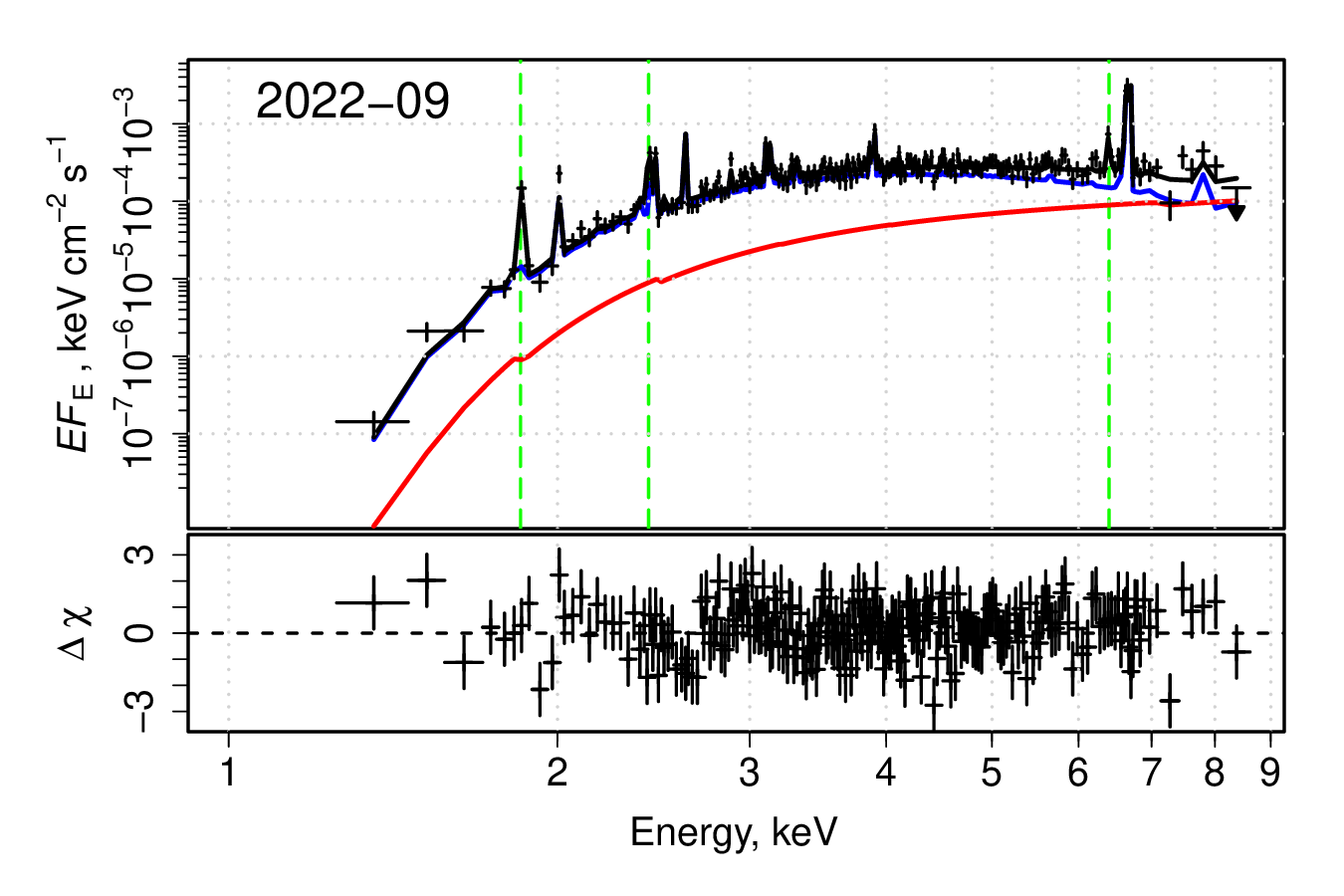}
    \caption{Chandra 2022 X-ray spectra of the Arches cluster between 1 and 9 keV for three epochs in 2022. Best-fitting model (solid black line) is composed by a combination of an absorbed \texttt{apec} (blue) and power-law (red). Green dashed lines show positions of {\ka} lines of He-like Si at ${\sim}1.85$~keV, S at ${\sim}2.45$~keV and Fe at $6.4$~keV (fixed), respectively, added to the fitting procedure (see Table~\ref{tab:spec:apec}).}
    \label{fig:apec:spe:chandra}
\end{figure}

Next, we extend the fitting energy range to 2$-$10~keV (1$-$9~keV) for {\xmm} ({\chan}) data, and include Arches cluster thermal emission modeled as optically thin, ionization equilibrium plasma \citep[\texttt{apec},][]{2001ApJ...556L..91S}. The spectral model is defined as follows:

\begin{equation}
\label{eq:apec}
    tbabs \times const \times (apec + pegpwrlw)
\end{equation}

where \texttt{tbabs} denotes the Tuebingen–Boulder ISM absorption model \citep{2000ApJ...542..914W}, and \texttt{pegpwrlw} is now defined in 2$-$10~keV band. Based on the previous observations of the Arches cluster \citep[][T12]{2006MNRAS.371...38W}, we fixed the metallicity to $Z = 1.7 Z{\odot}$ throughout the paper. {\chan} 2022 spectrum shows clear excesses above the model at ${\sim}1.85$ and ${\sim}2.45$~keV, most likely corresponding to the {\ka} lines from He-like Si and S, respectively. To account for of these features observed in many studies of the Arches cluster \citep[T12,][]{2017MNRAS.468.2822K,2025MNRAS.540.3850H}, we added two Gaussian lines in the model. The best-fitting model parameters are listed in Table~\ref{tab:spec:apec}, while X-ray spectra are demonstrated in Figs.~\ref{fig:apec:spe:xmm}-\ref{fig:apec:spe:chandra}. The spectra of the Arches cluster in 2020 and 2022 are described by the model with acceptable fit statistics. The Arches thermal emission is described by plasma temperature $kT\simeq2$~keV and non-thermal excess is characterized by a hard continuum component. The total  2$-$10~keV flux is dominated by the low temperature component, and about one third of the observed flux can be attributed to the power-law. The photon index $\Gamma$ measured with {\xmm} in 2020 is determined with range [$-0.8 \dots 1.5$]. {\chan} 2022 non-thermal component is described with power-law slope $\Gamma\approx1.5$ within range [$0.6 \dots 2.6$], which is consistent with $\Gamma=1.1\pm0.7$ in the Arches cluster core obtained for the long {\xmm} observation in 2015 \citep{2017MNRAS.468.2822K}, when X-ray intensity of the surrounding molecular cloud was significantly decreased, compared to 2000-2010.

\begin{table*}
\centering
\caption{Best-fitting parameters of \texttt{apec} and power-law model (Eq.~\ref{eq:apec}). }
\label{tab:spec:apec}
\begin{tabular}{lllllll}
\hline
 &  & \multicolumn{2}{c}{{\xmm} 2020} & \multicolumn{2}{c}{{\chan} 2022} \\ 
Parameter & Units & ObsID & Value &Epoch& Value\\ 
\hline
$const$ & &0862470701 & 1.0 (fixed) &2022-03&1.0 (fixed)\\ 
$const$ & &0862470601 & $1.17\pm0.07$ &2022-07&$1.04\pm0.03$\\ 
$const$ & &0862470801 & $1.00\pm0.07$ &2022-09&$1.02\pm0.03$\\ 
$N_{\rm H}$  & $10^{22}$~cm$^{-2}$ && $12.5_{-1.2}^{+1.5}$ &&$11.7_{-0.6}^{+0.7}$\\
$kT$ & keV && $2.09\pm0.25$ &&$2.16_{-0.18}^{+0.26}$\\
$Z/Z_{\odot}$ & && 1.7 (fixed) &&1.7 (fixed)\\
$I_{\rm apec}$ & $10^{-3}$ && $1.69_{-0.32}^{+0.50}$ &&$1.77\pm0.35$\\
$\Gamma^{\rm pow}$ & &&  $0.5_{-1.3}^{+1.0}$  &&$1.54_{-0.95}^{+1.06}$ \\ 

$F_{\rm 2-10~keV}^{\rm pow}$ & $10^{-13}$~\ergscm &&  $1.33\pm0.63$ &&$2.24\pm0.65$\\ 
$F_{\rm 2-10~keV}^{\rm total}$ & $10^{-13}$~\ergscm &&  $4.91\pm0.18$ &&$5.94\pm0.13$\\ 

$E_{\rm 1.85~keV}$ & keV &&--&& $1.843_{-0.010}^{+0.016}$ \\  
$F_{\rm 1.85~keV}$ & $10^{-7}$~\phflux &&--&& $7.18_{-3.46}^{+3.68}$ \\  

$E_{\rm 2.45~keV}$ & keV &&--&& $2.424\pm0.022$ \\  
$F_{\rm 2.45~keV}$ & $10^{-6}$~\phflux &&--&& $1.11_{-0.28}^{+0.32}$ \\  

$E_{\rm 6.4~keV}$ & keV &&--&& 6.4 (fixed) \\  
$F_{\rm 6.4~keV}$ & $10^{-7}$~\phflux &&--&& $5.28\pm2.60$ \\  

$\chi^{2}_{\rm red}$/dof & && 0.97/164 &&1.03/650\\ 
\hline
\end{tabular}
\end{table*}

To assess the significance of the power-law component detected with {\xmm} 2020 ({\chan} 2022) data, we repeated the fitting procedure without  \texttt{pegpwrlw} term. As expected, the temperature of \texttt{apec} increased to a higher value of  $kT=2.6\pm0.3$~keV ($2.4\pm0.1$~keV) and the quality of the fit worsened to $\chi^{2}_{\rm red}$/dof = 1.05/166 (1.1/652). The corresponding F-statistic test provides $p$-value of $4\times10^{-4}$ ($4\times10^{-11}$), which indicates that the power-law improves the fit.

\begin{figure}
\includegraphics[width=\columnwidth]{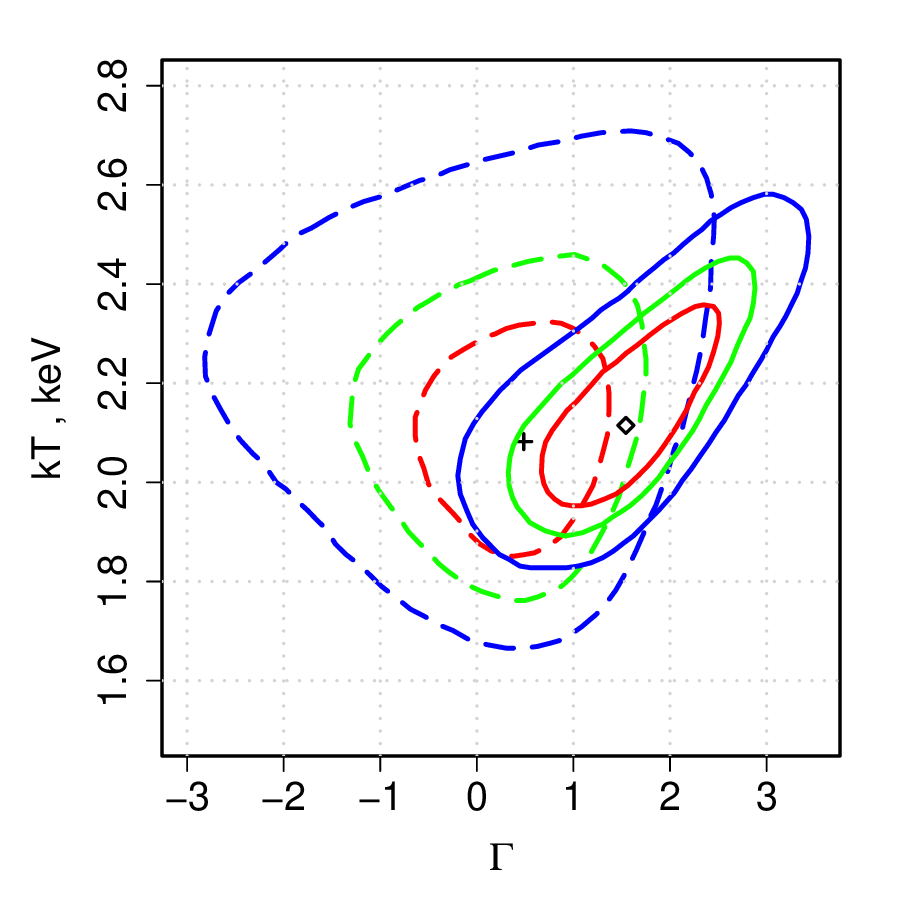}
    \caption{Two-dimensional $1\sigma$, $2\sigma$ and $3\sigma$ (red, green, blue colors) confidence contours for the slope $\Gamma$ of the non-thermal component and the temperature of the thermal plasma in the $2-10$~keV energy band for {\xmm} 2020 and Chandra 2022 data sets shown as dashed and solid lines, respectively. Cross and diamond show best-fitting values for {\xmm} 2020 and Chandra 2022 data, respectively.}
    \label{fig:cont}
\end{figure}

The searches for the non-thermal spectral components are usually complicated by the observed degeneracy between the temperature of the thermal plasma and the slope and/or normalization of the non-thermal power-law\footnote{It should also be noted that in some works, the authors use an additional thermal emission model with higher temperature, instead of a power law. This is quite reasonable, since the dense core of the Arches cluster contains Wolf-Rayet type stars and a large number of massive stars of early spectral classes, which allows us to expect the presence of plasma with higher temperatures.} \citep[e.g.][]{2022AstL...48..636K}. In Fig.~\ref{fig:cont} we plot confidence contours for the power-law slope $\Gamma$ and temperature of the thermal plasma, which shows the range of power-law slopes and corresponding $kT$ values, acceptable by the data. To better constrain spectral components and break the degeneracy, the broad-band energy coverage is needed, especially above 10 keV.

\section{Discussion}
\label{sec:discussion}

Deep {\chan} observations of the Arches cluster in 2022 provide higher statistics above 7~keV, compared to {\xmm} data in 2020, which allows us to better constrain the photon index of the power-law component ($\Gamma\approx1.5$ in range [$0.6 \dots 2.6$]). This result is in line with detection of the non-thermal X-ray emission of the rich stellar cluster RMC~136 located in the starforming Doradus 30 complex \citep{1960MNRAS.121..337F} of the LMC by \cite{2022A&A...661A..37S} using eROSITA/SRG observations. The young (1.0-2.5 Myr) star cluster RMC~136 hosts a rich population of massive stars with strong stellar winds \citep{2022A&A...663A..36B}, which makes it similar to the Arches cluster. With eROSITA/SRG, the X-ray spectrum of RMC~136 is well described with thermal emission from a non-equilibrium ionisation plasma ($kT>1$~keV) and non-thermal powerlaw with $\Gamma=1.3$, indicating a hard X-ray spectrum, similar to the Arches cluster, as shown in the current work. This adds evidence to similar mechanism of particle acceleration in the shocks of the winds of massive stars in compact stellar clusters. 

\begin{figure}
	\includegraphics[width=\columnwidth]{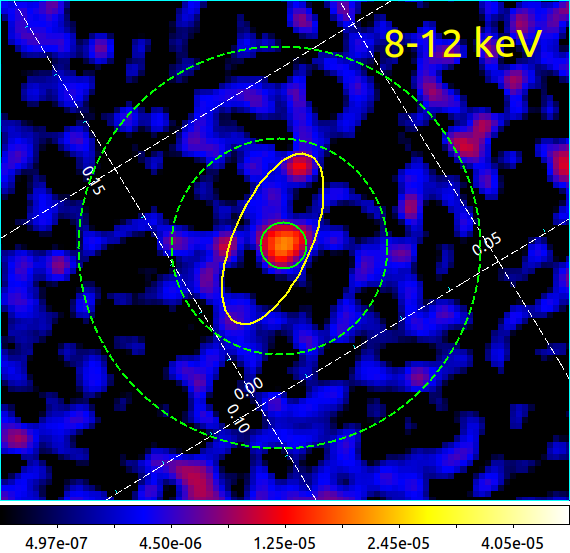}
    \caption{{\xmm}/EPIC exposure$-$ and background$-$ corrected map of the Arches cluster region in 8$-$12~keV range. For reference see Fig~\ref{fig:lines:map}.}
    \label{fig:cont:map}
\end{figure}

To constrain spatial extent of the Arches non-thermal emission in 2020, we constructed  MOS1, MOS2 and pn combined sky image of the cluster in hard 8$-$12~keV band, as described in Sect.~\ref{sec:xmm}. As seen from the resulted sky mosaic (Fig.~\ref{fig:cont:map}), the emission  is well localized within the cluster core. According to the best-fitting spectral model (Eq.~\ref{eq:apec}), the total flux is $F_{\rm 8-12~keV}=(1.1\pm0.3)\times 10^{-13}$~\ergscm, with relative contribution of \texttt{apec} and power-law components at the level of 26\% and 74\%, respectively.

% apec (2.854e-14 ergs/cm^2/s) range (8.0000 - 12.000 keV) 0.26
% pow (8.1385e-14 ergs/cm^2/s) range (8.0000 - 12.000 keV) 0.74
%
% text variant for 8-10 keV
%
%As seen from the resulted sky image in Fig.~\ref{fig:cont:map}, the emission with total flux of $F_{\rm 8-10~keV}=(5.9\pm0.8)\times 10^{-14}$~\ergscm\ is well localized within the cluster core. According to best-fitting spectral model (Eq.~\ref{eq:apec}), \texttt{apec} and power-law components constitute 36\% and 64\% in total 8$-$10~keV flux, respectively.
% apec (2.1085e-14 ergs/cm^2/s) range (8.0000 - 10.000 keV) 0.36
% pow (3.7602e-14 ergs/cm^2/s) range (8.0000 - 10.000 keV) 0.64

The morphology of diffuse non-thermal emission is different from that was reported in Westerlund 1 \citep{2006ApJ...650..203M} and Westerlund 2 \citep{2023MNRAS.525.1553B} where the emission apparently extended well beyond the cluster core. This difference can be understood, however, if the non-thermal emission is produced by the synchrotron radiation of very high energy electrons accelerated in the cluster core. In this case the extent of the synchrotron halo around the cluster would decrease with the increasing of the magnitude of the ambient magnetic filed, surrounding a cluster. The mG range  magnitude magnetic fields in the Galactic Center vicinity \citep[see e.g.][]{2024MNRAS.531.5012A} are substantially higher than that expected at the other Galactic locations of the clusters and this reduce the size of the synchrotron halo of the Arches cluster. The localized morphology of the non-thermal emission  helps us to distinguish it from the other non-thermal components present in the Galactic Center region.    

\section{Summary and conclusions}
\label{sec:summary}

In this work we characterize the X-ray emission of the Arches stellar cluster in a view of separating its thermal and non-thermal emission components. Previous studies have been concentrating on the reflected emission of the nearby molecular cloud. This emission, traced by the bright Fe {\ka} 6.4~keV line, prevented detailed investigation of the intrinsic cluster's non-thermal emission, usually attributed to synchrotron or inverse Compton scattering of relativistic electrons.  However, it should be noted that the compact core of the Arches cluster contains more than ten Wolf-Rayet stars and about one hundred massive stars of early spectral classes. This allows us to expect the presence of plasma with a temperature of 3$-$4~keV in certain regions of the cluster core. Therefore, in order to obtain a complete picture, it is necessary to clarify the spectral shape of the cluster's emission at energies above 10 keV.

After significant decrease of the reflected emission, observed after 2015, we got a chance to observe Arches cluster in the cleanest environment. Using available deep {\xmm} and {\chan} observations of the cluster in 2020 and 2022, respectively, we demonstrated that Fe {\ka} 6.4~keV line is not significantly detected within the cluster, providing evidence that contribution from the reflected emission of the molecular cloud is not significant. We showed that the overall X-ray spectral shape of the Arches cluster contains a soft, thermal component as well as a hard, non-thermal component. The spatial morphology of the latter is well localized in the cluster core region, and not extended beyond it. 

%To better constrain the Arches cluster non-thermal emission in the future studies, we need to break the spectral degeneracy with thermal plasma, and the high-energy data above 10 keV are especially needed.

\section*{Acknowledgments}
RK acknowledges support from the Russian Science Foundation (grant no. 24-22-00212). MC acknowledges financial support from the Centre National d’Etudes Spatiales (CNES).

\bibliographystyle{astl}
\bibliography{biblio} 

%================================================

%{\it Translated by V. Astakhov}

{\it Latex style was created by R. Burenin}

% Don't change these lines
%\bsp    % typesetting comment
\label{lastpage}

\end{document}